# Sensor Networks in Healthcare – Ensuring Confidentiality and User Anonymity in WBAN

Sayed Ashraf Mamun
*Rochester Institute of Technology*
sam7753@rit.edu

## Abstract

*Wireless body area network(WBAN) is becoming more popular in recent years. Security and privacy of this network is the major concern for the researchers. Most of the WBAN sensors currently available in the market uses Bluetooth Low Energy (BLE) protocol for connection. But the BLE protocol has an inherent security flaw due to its weak pairing protocol which can be easily be exploited by an attacker. In this paper, a different pairing protocol based on Lightweight Anonymous Authentication Protocol (LWAA) is proposed. Proposed protocol has a 3.8% power overhead over traditional BLE. But this is a very small price to pay if we consider all the benefits it brings to the system.*

## 1. Introduction

Wireless Sensor Network (WSN) is currently a hot research topic. WSNs are being used in more and more fields every day. Among these fields, one of the most important and critical fields is medical and health-care field. The life expectancy of an average person is increasing due to the medical advancement, improvement in nourishment and professional monitoring. Due to the proper monitoring and taking timely measures, health-care providers can take many preventive measures rather than doing treatment after any significant health issue occur. Many health monitoring systems like blood-pressure sensor, heart-rate monitoring sensor, body temperature sensor, ECG monitoring sensor, EEG sensor, movement and fitness activity sensor are already in use for a while. These sensors can be wearable on the body surface or implanted inside of the body. Collection of the sensed and measured data is most commonly done using some wireless network. But traditional and conventional wireless links and methods are not suitable for these small devices because of extreme resource constraints, mainly low battery life and low computational power. The network for these resource-constrained sensors which might or might not be inside of a human body are commonly called Wireless Body Area Networks (WBAN). As most other wireless networks, security and privacy is a burning issue here. There should be some kind of mechanism through which the attacker can be prevented from getting the information from the network or manipulate the data available in the network. In simple words, data confidentiality, integrity and user authentication should be ensured in this network. Moreover, maintaining the user anonymity is becoming almost an essential feature in current days. As the network and the devices are resource constrained, conventional measures cannot be taken to ensure data confidentiality and integrity. There are few security and privacy solutions available for WBAN architecture. Among them, Bluetooth Low Energy (BLE) is the most commonly used for WBANs. But there is a severe vulnerability in the pairing protocol of BLE. How this issue can be addressed, will be explored in this paper.

## 2. Literature review

The Idea of WBAN was first proposed by Zimmerman way back in 1996 [1]. But recently, due to the integration with the Internet of Things (IoT), WBAN has again become a hot topic for research as well as for the industry. WBAN is constructed with some resource-constrained sensors, which are placed on or inside of different parts of a human body. The measurements form these sensors can be monitored remotely. WBANs are most commonly used in the healthcare system. They provide a more accurate data about the patient to the healthcare provider because the data collected are from the normal activities in the natural environment of the patient, not from a controlled environment.

Sensors used in the health-care system can roughly be divided into two types- sensors which monitor vital status and sensors which are used for healthcare surveillance [2]. The vital status monitoring includes

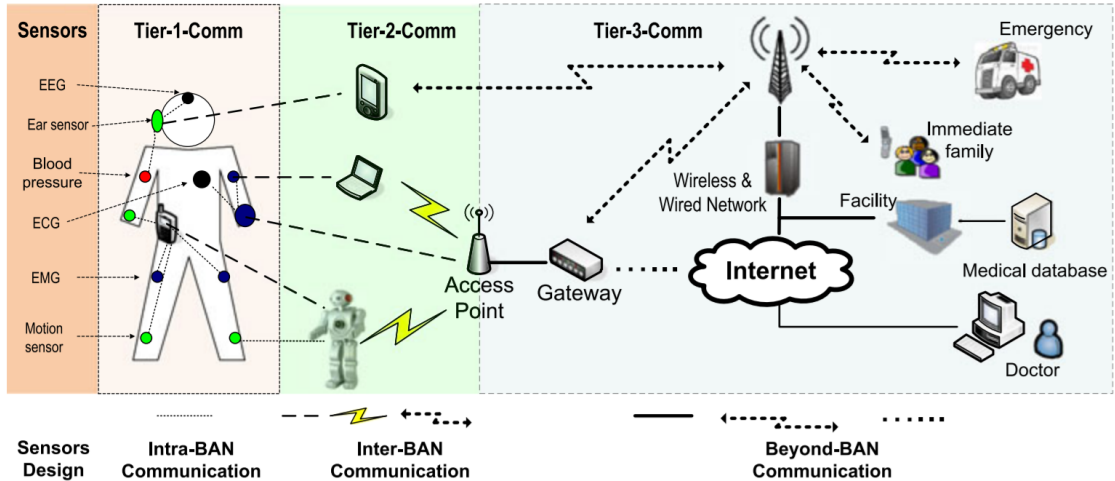
**Figure 1.** A three-tier architecture based on a BAN communications system [7]

patient's status monitoring inside a hospital, detection of epilepsy seizure etc. which requires immediate attention. On the other hand, healthcare surveillance can include elderly monitoring, temperature monitoring, ECG monitoring, EEG monitoring, which can be used to observe the long-term condition of a patient and might not need immediate attention.

The WBAN should be able to guarantee the followings stringent security and privacy requirements– security, data integrity and most importantly, confidentiality of the user's health record all the time [3]. The data sensed by the body network is very sensitive in respect to two aspects. Firstly, the data will shape the future diagnosis of the patient. Secondly, the privacy is an extremely important issue for patient's medical data due to the American Health Insurance Portability and Accountability Act (HIPAA). Sometimes, WBANs are also expected to provide data freshness, availability of the network, secure management, dependability, secure localization, accountability, flexibility and support privacy rules and compliance requirement [4].

Ensuring security of a sensor network is extremely important. In the worst case, a security breach can lead to the death of a patient [5] if the sensors are controlling some vital medical instruments like heart pace-maker or lungs ventilator. On the other hand, a business executive wearing a heart-rate or ECG monitoring sensor is practically transmitting a signal of a lie-detector. If the rival party can get hold of this signal, it can easily be used as an unfair advantage against that executive. There are many kinds of attacks that can happen to a WBAN apart from just trying to get an unauthorized access to the network. This includes jamming the wireless channel, physical tampering of the sensor, denial of service (DoS) [6], side channel attack, sinkhole attack, Sybil attack, wormhole attack etc.

A general architecture of WBAN centric healthcare system is consists of 3-Tier communication design [7] as shown in Figure 1. The 3-Tiers are intra-WBAN communication, inter-WBAN communication and finally, beyond-WBAN communication. Most of the research seen in the literature is seen to be limited in one of these three tiers. Different security aspects are required for each Tier. In this research proposal, only the Tier-1 portion of the WBAN architecture will be explored. Intra-WBAN communication is normally limited within 2 meters of a human body. Intra-WBAN communication can be further divided into 2 types- sensor-to-sensor communication, sensor-to-portable station (sink) communication. But in most of the literature, it is seen that the sensors in a WBAN communicate directly with the sink in a star topology, having the sink at the center of the star.

There are few wireless standards available for WSNs [8]. Wireless LAN (IEEE 802.11- WLAN) is the well-defined and accepted standard for home and office WiFi wireless network. Then there is IEEE 802.15.1 standard which is known as the Bluetooth protocol. It has three class of operation with a range of 100 meters, 10 meters and 1 meter respectively. But in a resource-constrained environment, it becomes impractical to use these standards as it is. Then came another standard namely IEEE 803.15.4, tailored for very low power wireless devices. It is also known as ZigBee. To standardize the WBAN communication, in 2007, Task group 6, which is also known as IEEE 802.15.6 was formed. It defines a Medium Access Control (MAC) which supports different Physical layers [9]. It offers 3 levels of security: unsecured communications, authentication only and finally authentication and encryption. There are also few other solutions available in literature- IrDA, NFC, ANT etc. In the following sections different available protocols for WBAN will be discussed briefly.

## 3. Threat model

For this paper, the following 3 threat/attack on the network has been considered:
i) Eavesdropper: They can try to eavesdrop on the wireless signals transmitted by the user's device.
ii) Man-in-the-middle(MITM): Here the attacker secretly relays or try to alters the communication between two parties, creating an illusion to the parties that they are talking with each other.
iii) Surveillor: They will try to identify the user by listening to his/her device and try to correlate with previously captured data.

## 4. WBAN communication protocols

In this section, few of the possible communication protocols those are currently in existence are discussed.

### 4.1. Bluetooth low energy (BLE)

One of the recently developed standard Bluetooth Low Energy (BLE) [10] has many lucrative features that can be utilized for WBANs. BLE is developed for short-range applications and consumes very low power. BLE is a wireless communication protocol adopted in many fields of our daily life. Applications of BLE are ranging from fitness, healthcare, entertainment to home security and many more. It is used in most of the currently developed wearable devices available in the market. It is a modified version of the traditional Bluetooth protocol. It uses the same 2.4 GHz wireless band radio frequencies spectrum as traditional Bluetooth. But BLE has 40 channels with 2 MHz channel spacing. Contrasting to the traditional Bluetooth's (Bluetooth BR/EDR) ability to connect with a maximum of 7 different connections, BLE can theoretically connect to an unlimited number of devices. Typical range of operation of BLE is within 50 meters and maximum transmission power is 10mW.

BLE has a stack layer Generic Access Profile (GAP) which decides the topology of the network. The device which initiates the connection process, normally act as the GAP central. Once two devices are connected, they will initiate a paring process where they exchange information to establish an encrypted connection.

BLE introduces many security and privacy issues in the network. The main security issues involved with the BLE protocol are man in the middle (MITM) attack, eavesdropping, and identity tracking. BLE already has support for 128-bit AES encryption with Counter Mode CBC-MAC (AES-CCM) to ensure the confidentiality and integrity of the data. BLE uses small data frames of 31 Bytes as shown in Figure 2, which makes any asymmetric encryption very hard and inefficient. But the main weakness of the BLE is the pairing stage. It has been demonstrated in the literature [11] that the entire security of the BLE network can be compromised by listening or overhearing only the connection establishment phase and capturing the required keys. The attacker only needs at maximum 20 pairing attempts captured to successfully break into the system. From every failed attempt, the attacker will get an incremental feedback of the security codes.

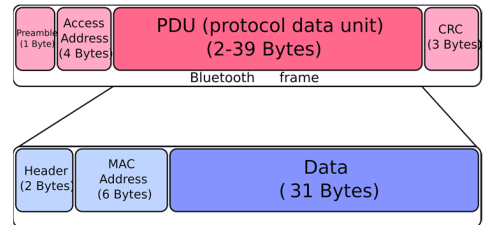

**Figure 2.** Bluetooth low energy Data Packet [10]

### 4.2. CryptoCop

To address the pairing and key exchange issue of the BLE, in 2016 a new protocol is proposed namely CryptoCoP [12]. In this protocol, the authors proposed that the key negotiation over the wireless channel is not required at all. They proposed the model for wearable devices only. As the wearable devices are needed to be charged periodically through cables, the authors envisioned that the keys can be exchanged during that time. By implementing the CryptoCoP, devices can get rid of the computationally expensive asymmetric encryption altogether. This helps in reduction of computational cost as well as radio transmission cost as there is no need for sending the long asymmetric key of 1024 bits or 2048 bits through the wireless. To ensure the data confidentiality in CryptoCoP, AES is used in Counter (CTR) mode. To make the system self-synchronized, an updated counter value is sent with the header. To ensure a second layer of security, keys are periodically rotated. To maintain the anonymity, CryptoCoP introduces a mechanism to be anonymous to a Surveillor, but still be able to pair-up with previously connected devices (i.e. sink). But one of the major disadvantages of CryptoCoP is that it will not work for implanted sensor inside of the human body at all as there is no way to connect cable to those sensors.

### 4.3. LIRA

A different protocol named LIRA [13] has been proposed with a different way to address the key exchange for the wearable body sensors. LIRA stands for *light* channel for sensor *initialization* and *radio* channel for *authentication*. In that paper, the authors proposed to use a separate light source unit to transmits secret keys over a protected visible light channel to the body control unit which is essentially the sink. The major disadvantage of this architecture is that it requires clear line-of-sight (LoS) between the sensor and the sink to work properly which might not be practical for wearable sensors, and impossible for implanted sensors.

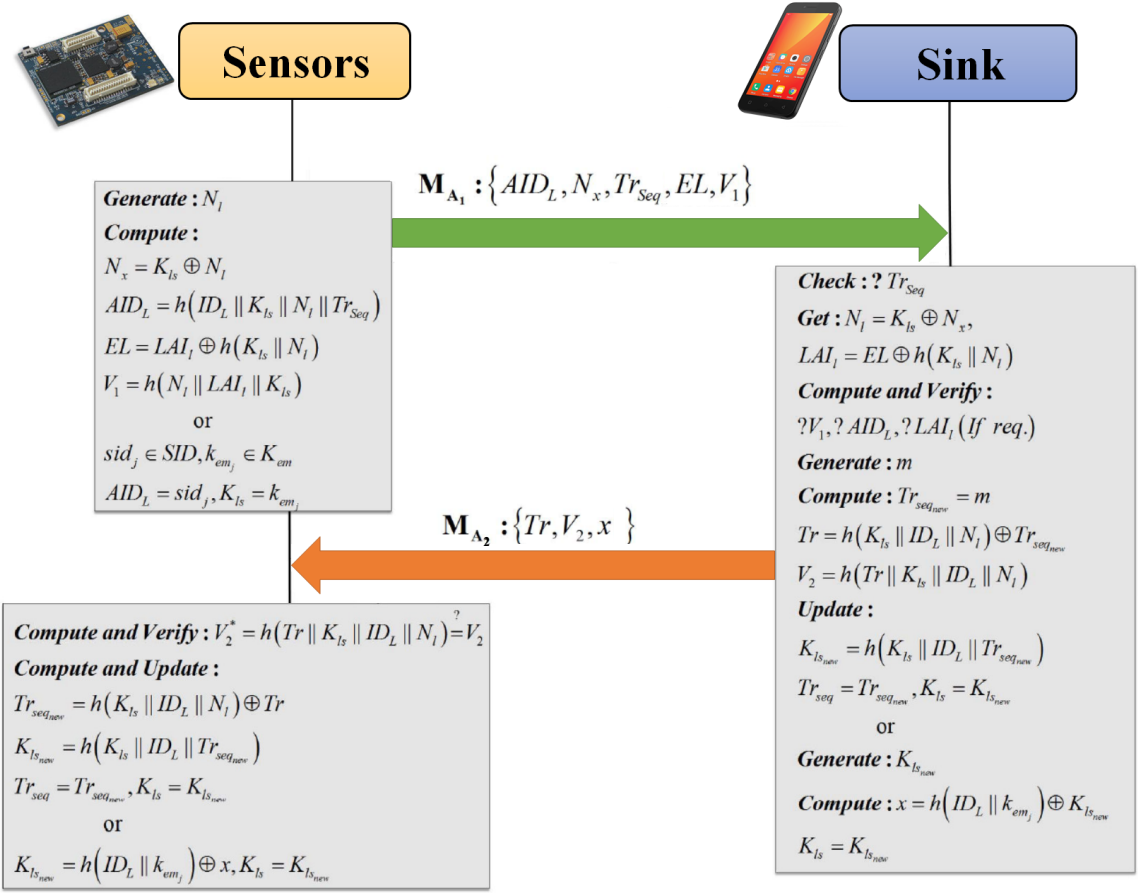

**Figure 3.** Lightweight anonymous authentication protocol

### 4.4. ZigBee

Zigbee [14] is an IEEE 802.15.4-based specification for a high-level communication protocol which is used to create small, low-power network. It is currently being used mainly for home monitoring devices and medical devices. The design goal for ZigBee is to establish low-power, low data rate, and close proximity ad hoc network.

Zigbee has the ability to deliver low-latency communication. To reduce the device cost, Zigbee chips are often integrated with microcontrollers and radios. Similar to BLE, ZigBee also operates at 2.4GHz band. It also supports few other frequency bands within the range of 700-900MHz. Data rate supported by ZigBee varies from 20kbps to 250kbps. ZigBee is built on the top of IEEE 802.15.4. It adapts the Physical and Link layers from the IEEE 802.15.4 standard. ZigBee uses 3 types of keys for the security purpose- link key, network key, and master key. But compared to the BLE protocol, the overhead for the ZigBee network is much higher and hence the power consumption is also high. This is due to the fact that the packet size for ZigBee is much higher than the BLE. In ZigBee, devices can work in two different mode to reduce the power consumption- full function device (FFD) and reduced function device (RFD).

There is another enhanced version of ZigBee came out in 2007 which is marketed as ZigBee Pro.

### 5. Proposed solution

In this section, a solution is proposed to alleviate the security problem associated with the pairing mechanism of the BLE protocol.

**Table 1.** Acronyms with definition

| Symbol | Definition |
|---|---|
| $ID_L$ | Identity of the sensor |
| $AID_L$ | One-time alias identity of sensor |
| $SID$ | Shadow identity of the sensor |
| $K_{ls}$ | Shared key between the sensor & sink |
| $K_{lsNew}$ | New shared key between the sensor & sink |
| $K_{em}$ | Shared emergency key between the sensor & sink |
| $Tr_{Seq}$ | Track sequence number |
| $Tr$ | New track sequence number |
| $LAI$ | Location area identifier |
| $h(.)$ | One-way hash function |

### 5.1. Assumptions

While proposing the alternate solution for the BLE pairing protocol, the following assumption were considered:

i) The wireless channel between the sensor and the sink is reliable.
ii) The sink is not a resource constrained device.
iii) Devices only have the ability to communicate with the sink. No inter-sensor communication is possible.
iv) Sensors always have enough data available to periodically send them to the sink.
V) Sensors are at a same distance from the sink. The transmitting power of every antenna is equal.

### 5.2. Lightweight anonymous authentication protocol

Instead of the prescribed pairing mechanism of the BLE protocol, in this paper we propose to use the Lightweight Anonymous Authentication Protocol [15] as the pairing and the key exchange mechanism. In [15], the authors proposed to use the Lightweight Anonymous Authentication (LWAA) protocol for the connection between the server and the sink. But in this paper, the LWAA protocol is proposed to be used for the connection between the sensor and the sink. There are two different connection phases of this protocol. The first phase is called the *registration* phase. For this phase, a secured channel is required and it is required to be done only once. It is envisioned that, it can be done before implementing the sensors on/inside the human body in a secure location. In the *registration* phase, sensor sends its identity number $ID_L$ to the sink through the secured channel. When the sink receives the $ID_L$, it calculates $K_{ls}= h(ID_L || N_s) \oplus ID_L$, where $N_s$ is a random number generated by the sink, and $h(.)$ is a hash function. The length of all the numbers and the output of the hash function, all are considered to have a length of 128bits each. After validating the sensor for the first and only time through $ID_L$, the sink sends $K_{ls}$, shadow ID, emergency key $K_{em}$ and the hash function $h(.)$ to the sensor.

The second phase is considered as the main lightweight anonymous authentication protocol. The entire process is shown in Figure 3. All the acronyms used in this paper are described in Table 1. In brief, the sensor sends a message $M_{A1}$: {$AID_L$, $N_x$, $Tr_{Seq}$, $EL$, $V_1$} to the sink. After validating the $M_{A1}$ message, the sink sends back another message $M_{A2}$: {$Tr$, $V_2$, $x$} to the sensor. The sensor and the sink authenticate each other by $M_{A2}$ and $M_{A1}$ respectively. Finally, the $K_{lsNew}$ which is generated by the sink, will be used by the sensor as the encryption key for AES.

### 5.3. Simulation

The LWAA protocol was implemented in MATLAB to check whether the protocol works as described in [15] or not. A partial simulation of the network was done only consisting of the anonymous authentication process to find the computational and communication overhead of the LWAA protocol. This was done to do a comparative analysis of the proposed method with some already existing protocols, namely- traditional BLE, CryptoCoP, and ZigBee. To measure the power consumption of an individual sensor, it is assumed that, every sensor has to send 3 packet worth of information to the sink in every 5-second interval and go to sleep mode. After waking up from the sleep stage, sensors need to pair up with the sink every time. Computation cost per-cycle is considered to be 0.00354 μJ according to [16]. To have a fair comparison between the protocols, for ZigBee, transmission (Tx) and receive (Rx) power for only 2.4GHz band was considered as all other also work in that same band. Per-bit Tx/Rx power consumptions for all the protocols were adapted from [17]. As all the discussed protocols use AES-128 for encryption, the same cost of 38 μJ/block for encryption is considered for all of their power consumption according to [16]. The detail calculation of power consumption is shown in Table 2.

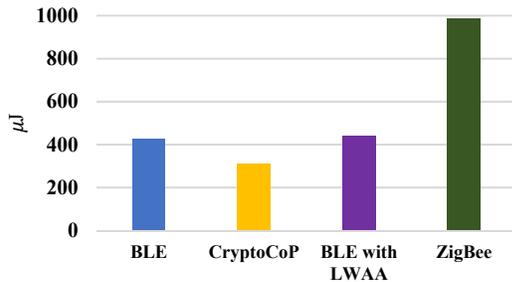

**Figure 4.** Power consumption for different protocol (for every 5 second interval)

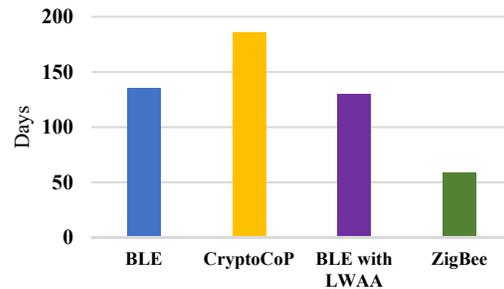

**Figure 5.** Average sensor life for different protocol

Table 2. Power Consumption Breakdown

| Protocols | Computation cost For Pairing | encryption Cost | For Pairing/Authentication | | | | For Data packets/frames | | | | Total packet cost (for 3 packets) | Total power | Life |
|---|---|---|---|---|---|---|---|---|---|---|---|---|---|
| | | | number of bits Tx | total Tx cost | number of bits Rx | total Rx cost | number of bits Tx | total Tx cost | number of bits Rx | total Rx cost | | | |
| units | μJ | μJ/ block | | μJ | | μJ | | μJ | | μJ | μJ | μJ | days |
| **BLE** | 17.7 | 38 | 168 | 35.11 | 280 | 63.28 | 376 | 78.584 | 56 | 12.656 | 273.72 | 427.81 | 135 |
| **CryptoCoP** | 0 | 38 | 0 | 0 | 0 | 0 | 376 | 78.584 | 56 | 12.656 | 273.72 | 311.72 | 185 |
| **BLE LWAA** | 50.82 | 38 | 288 | 60.19 | 96 | 21.69 | 376 | 78.584 | 56 | 12.656 | 273.72 | 444.43 | 130 |
| **ZigBee** | 14.16 | 38 | 1024 | 214.01 | 88 | 19.89 | 1024 | 214.02 | 88 | 19.888 | 701.71 | 987.78 | 58 |

### 5.4. Results and analysis

The comparative results between different protocols are shown in Figure 4. The same result is translated into the sensor lifetime in Figure 5. The battery power for all sensors is assumed to be 1000mAh. From the Figure 4 and Figure 5, it is seen the performance of CryptoCoP is the best considered the power consumption or sensor life. But as discussed in Section 4.2, this method is not suitable for all kind of WBANs- precisely implanted sensor devices. Next best performing protocol is found to be conventional BLE. But this protocol also comes with its disadvantages as described in Section 4.1. ZigBee performs worst among the four compared protocols here as the communication overhead of ZigBee protocol is the highest.

The main observation from the results is that, the proposed protocol (BLE with LWAA) performs comparable to the traditional BLE, only having a 3.8% increase in power overhead. But this small increase in power consumption should not pose any real problem as it comes with the added features like- mutual authentication, anonymity, secure key exchange etc. Proposed protocol has minimal effect on battery life.

### 6. Conclusion

In this paper, a new pairing protocol is proposed for BLE protocol for WBAN networks. By implementing Lightweight Anonymous Authentication (LWAA) protocol over BLE, the vulnerability of the traditional BLE protocol's pairing can be alleviated and few added advantages like mutual authentication, anonymity can be achieved. The proposed protocol only adds about 3.8% power overhead over the conventional BLE protocol.

This number might be affected if there is some change in some of the assumptions considered in this paper. For example, it is considered that the wireless channel between the sensor and sink is very reliable, which might not always be true. If the channel is not reliable, the overhead number will certainly increase. Another assumption considered here is that all the sensors have always readily available data for sending to the sink, which also, might not always be true. The effect of these variations can be explored in the future.